# Recent advances in the combination of nonlinearity and exceptional points

Kai Bai,[1] Chen Lin,[1] Jia-Zheng Li,[2] Meng Xiao,[1, 3*]

[1]Key Laboratory of Artificial Micro- and Nano-structures of Ministry of Education and School of Physics and Technology, Wuhan University, Wuhan 430072, China

[2]Department of Physics and Center for Theory of Quantum Matter, University of Colorado, Boulder, CO 80309, USA

[3]Wuhan Institute of Quantum Technology, Wuhan 430206, China

Corresponding Email [*]: phmxiao@whu.edu.cn;

**ABSTRACT:**

The exotic physics emerging at singularities has long attracted intense theoretical and experimental attention. In non-Hermitian systems, exceptional points (EPs)—unique spectral singularities—have given rise to a host of intriguing wave phenomena and enabled a broad range of promising applications across diverse physical platforms. Recently, considerable effort has been devoted to combining nonlinearity with exceptional points (EPs) to enable flexible control, overcome the limitations of linear EPs, discover previously unexplored singularities, and reveal novel physical phenomena and application potentials. In this review, we provide a detailed overview of the interplay between nonlinearity and EPs, highlighting key developments such as noise suppression for enhanced sensing, emerging mechanisms for chiral-like state transfer, the realization of optical isolators in nonlinear EP systems, applications including wireless energy transfer and frequency comb generation, among others. We also offer a perspective on future research directions and opportunities in this rapidly evolving field.

**TABLE OF CONTENTS**

# I. INTRODUCTION

Non-Hermitian physics has emerged as a powerful framework for describing and engineering open classical and quantum systems that exhibit nonconservative phenomena, such as gain, loss, and non-Hermitian couplings [1-4]. Traditionally, the behavior of non-Hermitian systems has been inferred from their Hermitian counterparts by treating nonconservative elements as small perturbations, which has led to the distinct physics of non-Hermiticity being largely overlooked. Recent advances in the non-Hermitian systems in connection with exceptional points (EPs) [3,5,6] have revealed that nonconservative elements can fundamentally reshape system behavior, leading to effects that deviate significantly from those observed in Hermitian systems. An EP is defined as a spectral singularity at which two or more eigenvalues and their corresponding eigenvectors coalesce. After the physical existence of EPs was experimentally demonstrated in microwave cavities [7], EPs were subsequently experimentally observed in various classical and quantum systems [8,9,18-27,10,28-31,11-17], including microwave cavities [10], optical microcavities [11], coupled atom-cavity systems [12], photonic crystal slabs [13], exciton-polariton billiards [14], parity-time (PT) symmetric synthetic photonic lattices [15], acoustic systems [16], magnon–polaritons [8,21], quantum walk [23,24], superconducting circuit [25], atomic Bose-Einstein condensates [19], thermal atomic ensembles [20], single-spin systems[17], among others. A wide array of novel phenomena and potential applications related to EPs have been elucidated. For instance, EPs are responsible for amplification of a targeting signal [10,32] in microtoroid cavities, mode discrimination in multimode laser cavities [33,34], and pump-induced [35] or loss-induced [36] revival of lasing in coupled laser cavities. In addition, a dynamic loop near an EP in parameter space can lead to a chiral state transition, non-adiabatic state jumps, and topological mode conversion [37-40],which have been harnessed for topological energy transfer in optomechanical systems [37], polarization states conversion in an optical waveguide [38], chiral heat transport in a thermal system[27] performance enhancement of quantum devices [41,42], and on-chip optical devices such as optical isolators [43,44], and directional lasing [45].

On another front, nonlinearity is ubiquitous feature across a wide range of physical systems and plays a crucial role in governing complex dynamical behavior across optics [46-48], atomic gases [49,50], mechanical metamaterials [51–53], electronic circuits [54–56], and beyond [23,57–61]. Unlike linear systems, where superposition principle holds, nonlinear systems exhibit amplitude-dependent responses that enable soliton formation [62–65], harmonic generation [48,54], active tunability [55,66–72], genuine nonreciprocity [73–75], quantum light generation [76–78], and so on [79–87]. Notably, non-Hermiticity and nonlinearity are deeply intertwined. First, non-Hermitian components such as gain and loss are generally intensity-dependent, making them nonlinear; thus lasers, amplifiers, and saturable absorbers all exhibits both nonlinearity and non-Hermitian dynamics. Second, nonlinear optical effects such as the Kerr nonlinearity can couple wave components at different frequencies, allowing a strong pump wave to transfer energy to weaker probe waves. The effective operator governing the dynamics of each frequency component is intrinsically non-Hermitian, since the energy within that component is not conserved. Therefore, a fundamental question arises: What impacts do nonlinearities have on non-Hermitian systems exhibiting EPs, or vice versa?

Recently, considerable efforts have been devoted to exploring the synergy between EPs and nonlinearity, which has rapidly emerged as a vibrant research frontier [see Fig. 1]. This interplay has already given rise to a wealth of intriguing developments [23,44,91–100,67,101–110,68,111–118,71,72,74,88–90], as illustrated in Figs. 1(b–i), ranging from nonlinear tuning of non-Hermiticity associated with EPs [67,68,71,72,88,89,92], and higher-order exceptional singularities [93–96] to NEP-based sensing [97,98,122,99–103,119–121], the discovery of new singularities [98,104,105], chiral-like state transfer mechanisms [23,106,117,118], robust wireless power transfer [107–113], enhanced frequency combs[114], and nonreciprocal wave propagation [44,74,115,116].

In this Review, we systematically synthesize and compare representative recent studies arising from the integration of EPs and nonlinearity, with the overall structure schematically illustrated in Fig. 1. To begin, Sec. II presents a concise overview of the experimental realizations and associated applications of EPs across a variety of platforms (for more detailed discussions of linear EPs, we refer the reader to comprehensive reviews [3,123–128]), together with the mechanisms by which nonlinearity is incorporated into these systems [see Fig. 1(a)]. We then organize the core content of the Review around four unifying themes that capture the underlying physical mechanisms and conceptual advances enabled by the interplay between EPs and nonlinearity. In Section III, we discuss how the combination of nonlinearity and EPs enables flexible and real-time control of systems under consideration [see Fig. 1(b, c)]. Section IV explores how such interplay can overcome or mitigate the intrinsic limitations of linear EPs [see Fig. 1(d, e)]. Section V focuses on new singularities and mechanisms that emerge from the nonlinear–EP synergy [see Fig. 1(f, g)]. Section VI surveys novel physical phenomena and potential applications [see Fig. 1(i-h)]. Finally, Sec. VII provides an outlook on future opportunities and open challenges in this rapidly evolving field.

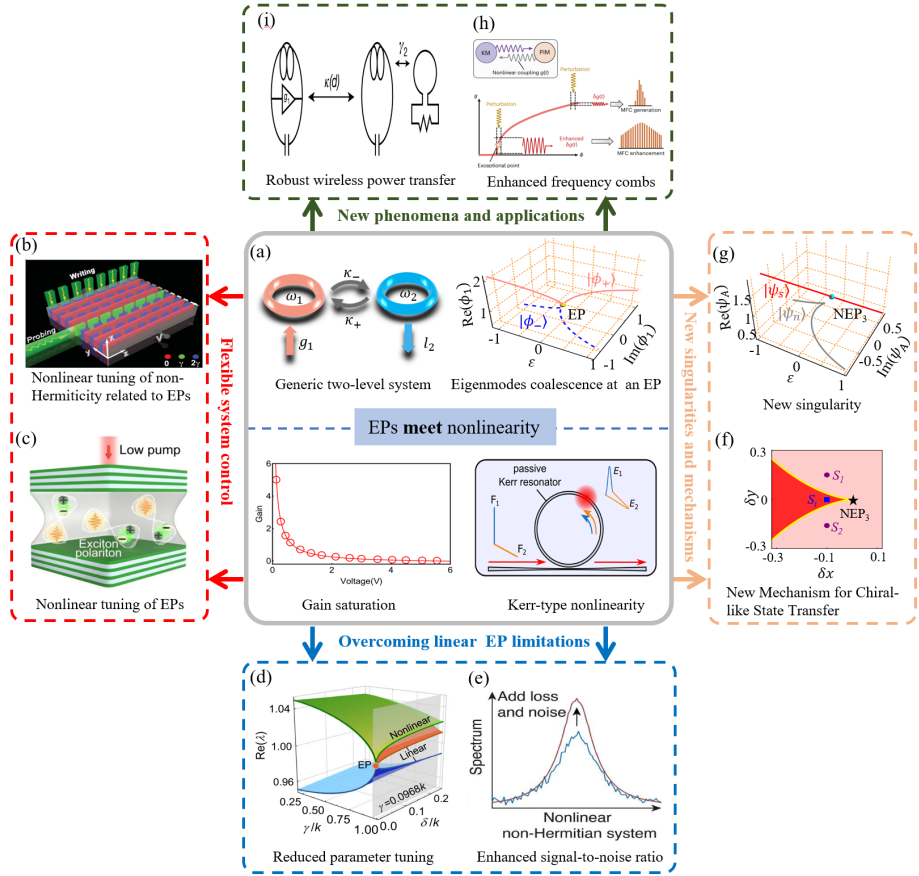

FIG.1. Structure of this review. (a) EPs meet nonlinearity: a generic non-Hermitian two-level system [98] with gain–loss contrast and asymmetric couplings, where the two eigenstates coalesce at an EP, together with typical nonlinear mechanisms such as gain saturation [94] and Kerr-type responses [129]. (b–i) The interplay between nonlinearity and EPs can be broadly organized into four functional categories. Flexible system control via nonlinearity, exemplified in (b) by nonlinear tuning of parity–time (PT) symmetry associated with EPs and non-Hermitian topological states [68], and in (c) by nonlinear tuning

that enables the emergence and observation of EPs [130]. Overcoming limitations of linear EP physics, illustrated in (d) by realizing higher-order exceptional singularities with fewer tuning parameters [99] and in (e) by enhancing the signal-to-noise ratio near nonlinear EPs [131]. New singularities and mechanisms, including the nonlinear chiral-like state transfer requiring minimal control parameters [118] (f) and the emergence of nonlinear exceptional points [98] (g). New phenomena and applications, such as enhanced frequency-comb generation [132] (h), and robust wireless power transfer [107] (i). The upper subpanel in panel (a) and panel (g) are reproduced with permission from Bai et al., Phys. Rev. Lett. 132(7), 073802 (2024). Copyright 2024 American Physical Society. The lower-left subpanel in panel (a) is reproduced from Bai et al., Natl. Sci. Rev. 10, nwac259 (2023); licensed under the Creative Commons Attribution (CC BY) license. The lower-right subpanel in panel (a) is reproduced from Xu et al., Nat. Commun. 12(1), 4023 (2021), licensed under the Creative Commons Attribution (CC BY) license. (b) Reproduced from Xia et al., Science 372(6537), 72–76 (2021). Copyright 2021 American Association for the Advancement of Science. (c) Reproduced from Opala et al., Optica 10(8), 1111-1117 (2023), licensed under the Creative Commons Attribution (CC BY) license. (d) Reproduced from Chen et al., Nat. Commun. 15(1), 9312 (2024), licensed under the Creative Commons Attribution (CC BY) license. (e) Reproduced from Li et al., Sci. Adv. 9(27), eadi0562 (2023). Copyright 2023 American Association for the Advancement of Science. (f) Reproduced from Bai et al., Nat. Commun. 16(1), 5844 (2025), licensed under the Creative Commons Attribution (CC BY) license. (h) Reproduced from Wang et al., Nat. Phys. 20(7), 1139–1144 (2024). Copyright 2024 Springer Nature. (i) Reproduced from Assawaworrarit et al., Nature 546(7658), 387–390 (2017). Copyright 2017 Springer Nature.

## II. BACKGROUND

### A. Linear exceptional points and their manifestations in non-Hermitian systems

EPs have been extensively investigated across a broad spectrum of classical and quantum systems. Figure 2 illustrates several representative examples. For open quantum systems, a rigorous treatment— such as employing the Lindblad master equation or the Heisenberg–Langevin formalism—is required to capture their dynamics. In this Review, however, we only focus on classical systems or quantum counterparts in the semiclassical limit, for which the dynamics can be effectively described by a non-Hermitian Hamiltonian.

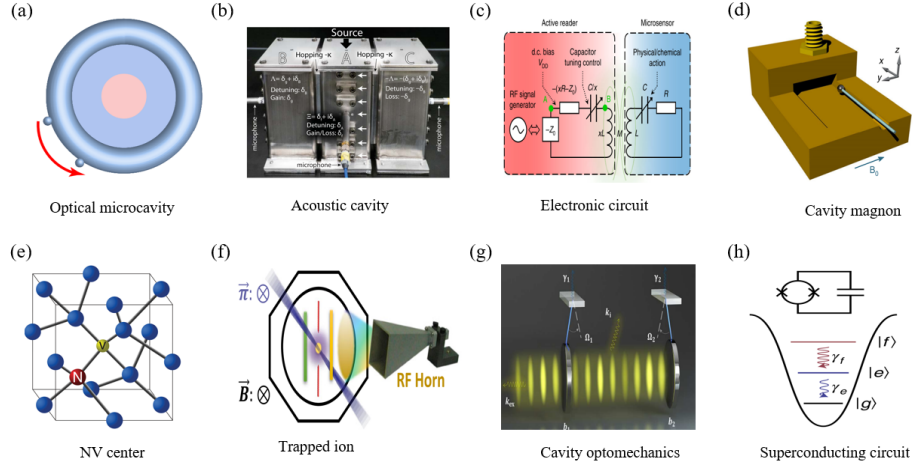

FIG. 2. Representative platforms for exploring EPs. EPs have been experimentally realized in a broad variety of classical and quantum systems, including optical microcavity [10] (a), acoustic cavity [16] (b), electronic circuit [30] (c), cavity magnon [21] (d), nitrogen–vacancy (NV) center in diamond [17] (e), trapped ion [29] (f), cavity optomechanics [133] (g), and superconducting circuit [18] (h). The dynamics of these diverse implementations can be effectively captured by a non-Hermitian Hamiltonian. (a) Reproduced from Chen et al., Nature 548(7666), 192–196 (2017). Copyright 2017 Springer Nature. (b) Reproduced from Tang et al., Science 370(6520), 1077–1080 (2020). Copyright 2020 American Association for the Advancement of Science. (c) Reproduced from Chen et al., Nat. Electron. 1(5), 297–304 (2018). Copyright 2018 Springer Nature. (d) Reproduced from Zhang et al., Nat. Commun. 8(1), 1386 (2017), licensed under the Creative Commons Attribution (CC BY) license. (e) Reproduced from Wu et al., Science 364(6443), 878–880 (2019). Copyright 2019 American Association for the Advancement of Science. (f) Reproduced from Ding et al., Phys. Rev. Lett. **126**(8), 083604 (2021). Copyright 2021 American Physical Society. (g) Reproduced from Li et al., Sci. Adv. 9(27), eadi0562 (2023). Copyright 2023 American Association for the Advancement of Science. (h) Reproduced from Naghiloo et al., Nat. Phys. 15(12), 1232–1236 (2019). Copyright 2019 Springer Nature.

We begin with a generic non-Hermitian two-level system [see Fig. 1(a)], which consists of two resonant modes with resonant frequencies $\omega_1$ and $\omega_2$, with one exhibiting linear gain $g_1$ and the other experiencing linear loss $l_2$. The coupling between them is non-Hermitian [134,135] and given by $\kappa_\pm = (\kappa_0 \pm \kappa_\varsigma) + i(\kappa_\nu \pm \kappa_f)$ with $\{\kappa_0, \kappa_\varsigma, \kappa_\nu, \kappa_f\} \in \mathbb{R}$. The dynamics of this system is governed by

$$i\frac{d}{dt}|\boldsymbol{\phi}^R\rangle = H_l|\boldsymbol{\phi}^R\rangle, \tag{1}$$

where $|\boldsymbol{\phi}^R\rangle \equiv (\phi_1, \phi_2)^T$ is the right eigenstate with the superscript $T$ indicating transpose, $\phi_1$ and $\phi_2$ represent the field amplitudes inside the two resonators. The corresponding tight-binding Hamiltonian $H$ is given by

$$H = \begin{pmatrix} \omega_1 + ig_1 & \kappa_+ \\ \kappa_- & \omega_2 - il_2 \end{pmatrix}. \tag{2}$$

For clarity, we set $\kappa_0 = 1$, as its value does not affect the discussion. An EP of $H_l$ occurs at $\kappa_\varsigma = \kappa_f = 0$, $g_1 = l_2 = \kappa_\nu = 1$, $\omega_1 = 0$, and $\omega_2 = 2$. we introduce an external perturbation parameter $\varepsilon$ imposed on EP along $\omega_1$. The eigenfrequencies of $H_l$ are then given by

$$\omega_\pm = \frac{2 + \varepsilon \pm \sqrt{\varepsilon^2 + (4i - 4)\varepsilon}}{2}, \tag{3}$$

with the corresponding eigenstates

$$|\boldsymbol{\phi}_\pm\rangle = \begin{pmatrix} \phi_{1,\pm} \\ \phi_{2,\pm} \end{pmatrix} = \begin{pmatrix} (i + (\frac{1}{4} - \frac{i}{4})(\varepsilon \pm \sqrt{\varepsilon^2 + (4i - 4)\varepsilon}) \\ 1 \end{pmatrix}. \tag{4}$$

For demonstration purposes, we set $\phi_{2,\pm} = 1$ ( $|\boldsymbol{\phi}_\pm\rangle$ is not normalized), so that $\phi_{1,\pm}$ alone characterizes the behavior of the eigenstates. The upper-right panel of Fig. 1(a) shows the evolution of $\phi_{1,\pm}$ as a function of the external perturbation $\varepsilon$. As indicated by the yellow dots, the two eigenmodes coalesce at the EP, reducing to a single state $(i, 1)^T$ with degenerate eigenfrequencies $\omega_0$.

Near the EP, the eigenfrequencies reduce to $\omega_{\pm} \approx \omega_0 \pm \sqrt{(i-1)}\sqrt{\varepsilon}$. This square-root dependence implies that even a small perturbation $\varepsilon$ can induce a large change in the eigenfrequencies [see e.g., Fig. 3(a)], which is often regarded as a key figure of merit for next-generation sensing technologies. By contrast, conventional sensors typically operate near other types of degeneracies, such as Dirac points (DPs), where the perturbation strength and frequency splitting exhibit only a linear relationship [136]. Consequently, theory predicts [137,138] that EP-based sensors can yield a much stronger response to small perturbations than their DP counterparts. This responsivity scales as $\varepsilon^{-1/2}$, approaching infinity as $\varepsilon \to 0$, which has been experimentally demonstrated [see Fig. 3(b)] in microtoroid cavities [10,139]. In addition, integrated microring resonators have been used to realize a third-order EP [see Fig. 3(c)], exhibiting a cube-root eigenfrequency splitting as the three degenerate modes are lifted by perturbations [32]. Since then, EP-based sensors have been implemented in diverse forms, ranging from single- or few-nanoparticle detection and temperature monitoring, to refractive index sensing and gyroscopes. Besides frequency measurements, the amplitude[140] and phase difference[119,120] can also serve as sensitivity indicators, enabling enhanced performance over a broader dynamic range. More comprehensive discussions about the enhancement of responsivity can be found in the relevant review articles [123–125,141].

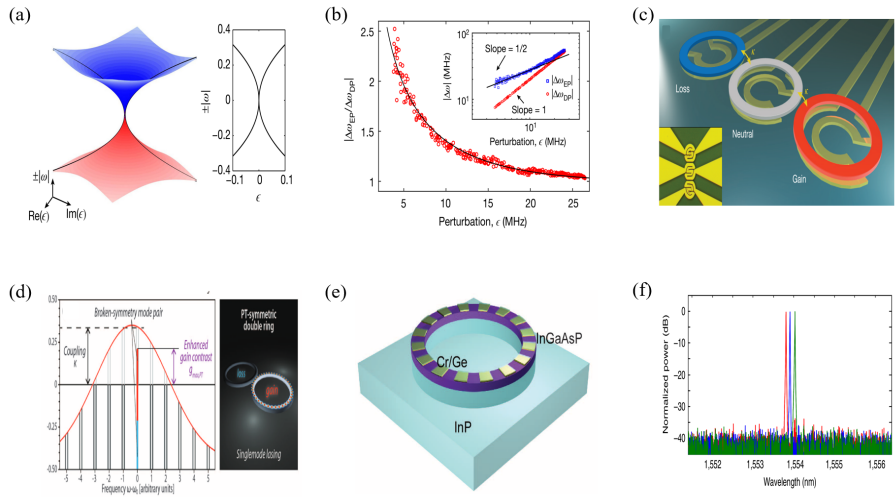

FIG. 3. Enhanced responsivity and related applications in sensing (a - c) and mode discrimination (d - f). (a) Complex frequency surfaces near an second-order EP, illustrating frequency splitting under a perturbation $\epsilon$ [10]. (b) Measured dependence of the enhanced complex-frequency splitting on the perturbation strength $\epsilon$ [10]. The inset shows the corresponding logarithmic plot, where the EP sensor exhibits a slope of 1/2 (black solid line) for sufficiently small perturbations, confirming the square-root behavior of a second-order EP. (c) Three coupled microring resonators realizing a third-order EP, where the perturbation-induced eigenfrequency splitting follows a cubic-root dependence [32]. (d) Schematic of a PT symmetric microring configuration used for mode discrimination [34]. Near an EP, a greatly amplified gain contrast can thus be obtained. (e) Single-mode laser via PT-symmetry breaking in single rings with active–passive gratings [33]. (f) Wavelength-tunable single-mode micro ring laser based on ring resonators [142]. (a, b) Reproduced from Chen et al., Nature 548(7666), 192–196 (2017). Copyright 2017 Springer Nature. (c) Reproduced from Hodaei et al., Nature 548(7666), 187–191 (2017). Copyright 2017 Springer Nature. (d) Reproduced from Hodaei et al., Science 346(6212), 975–978 (2014). Copyright 2014 American Association for the Advancement of Science. (e) Reproduced from Feng et al., Science 346(6212), 972–975 (2014). Copyright 2014 American Association for the Advancement of Science. (f) Reproduced from Liu et al., Nat. Commun. 8(1), 15389 (2017), licensed under the Creative Commons Attribution (CC BY) license.

Enhanced responsivity at the EP has also been exploited for mode discrimination in multimode laser cavities [143]. Generally, laser cavities support a large number of closely spaced modes because their dimensions are typically much larger than the optical wavelength and they exhibit a broad gain bandwidth [see the red line in Fig. 3(d)]. As a result, the output of such lasers often contains several modes simultaneously. Achieving single-mode operation in a given configuration has therefore been one of the primary goals of cavity design. In principle, any resonator with a spectrally nonuniform gain distribution can achieve single-mode operation by introducing a global loss that overcompensates the gain for all but one resonance, thereby eliminating the net gain of competing modes. However, in this regime the amplification cannot exceed the gain contrast $g_{max} = g_0 - g_1$, where $g_0$ denotes the gain of the principal mode and $g_1$ that of the next strongest competing mode. Clearly, this approach imposes severe constraints on the operating parameters, particularly in the case of broad gain windows and/or closely spaced resonator modes, where $g_{max}$ becomes very small. Interestingly, a large gain contrast can be achieved at the EP supported by parity–time (PT)-symmetric [144,145] systems [see Fig. 3(d)].

Specifically, in PT-symmetric arrangements the EP marks the transition from the exact to the broken PT phase, where the spectrum ceases to be entirely real and becomes partially complex. In this regime, all undesirable modes remain confined to the exact-PT phase, while the principal mode is driven into the broken-PT phase. Owing to the significantly enhanced spectral responsivity at the EP, a greatly amplified gain contrast can thus be obtained. This strategy has been implemented in coupled microring lasers [34] [Fig. 3(d)], in single rings incorporating active–passive gratings [33] [Fig. 3(e)], and in various other systems[146–149]. Furthermore, phase modulators in coupled rings can provide continuous wavelength tuning, yielding wavelength-tunable single-mode microring lasers [142] [Fig. 3(f)].

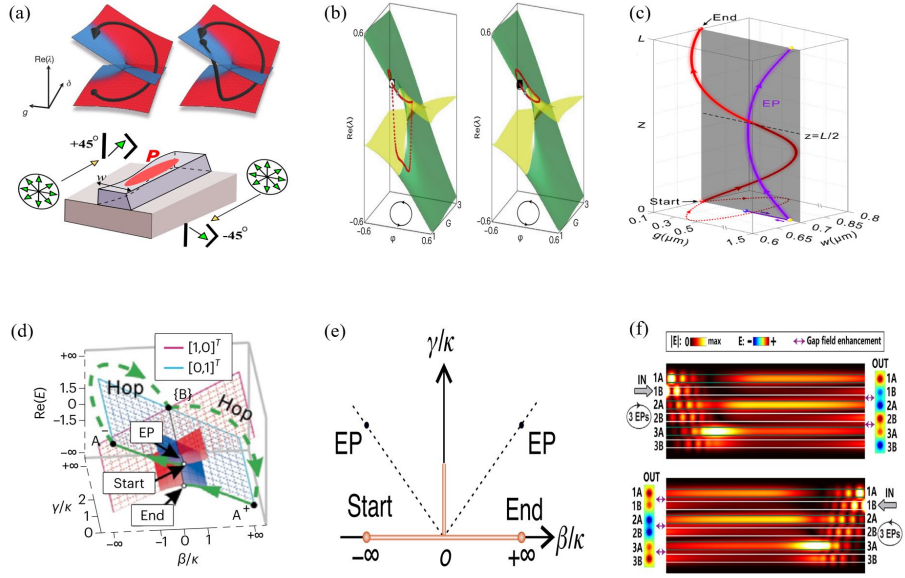

FIG. 4. Chiral state transfer and its applications. (a) The upper panel shows the eigenvalue surfaces near an EP [39]. Nonadiabatic transitions inevitably arise when slowly encircling an EP, giving rise to chiral state transfer (CST). The lower panel illustrates a CST-based application, an omnipolarizer [38] that forces the output polarization into a fixed direction regardless of the input polarization. (b) CST without encircling an EP [40]. (c) Efficient CST achieved by encircling moving EPs [150]. (d) Efficient CST via Hamiltonian jumps during encirclement [151]. (e) CST along an open trajectory without encircling EPs [152]. (f) Dynamical encircling of multiple EPs [153]. The upper subpanel in panel (a) is reproduced with permission from Doppler et al., Nature 537(7618), 76–79 (2016). Copyright 2016 Springer Nature. The lower subpanel in panel (a) is reproduced with permission from Hassan et al.,

Around EPs, the eigenvalue surfaces form self-intersecting Riemann sheets in parameter space [see the upper panel of Fig. 4(a)], giving rise to a rich non-Hermitian spectral topology [125,154,163,155–162]. To illustrate this, we first consider a Hermitian system. Let $\varepsilon \in \mathbb{R}^d$ denote the location in a $d$-dimensional parameter space. Hermiticity enforces that each eigenvalue $\omega(\varepsilon) \in \mathbb{R}$ remains confined to the real axis. As a result, when $\varepsilon$ is varied along a path $C_\varepsilon$, its image in the eigenvalue space, denoted $C_\omega^H$, forms a topologically trivial path on the real axis, precluding the emergence of nontrivial spectral topology [125]. In contrast, eigenvalues in non-Hermitian systems generally become complex-valued functions of $\varepsilon$, which expands the dimensionality of the complex-eigenvalue manifold $\mathcal{M}_\omega(\varepsilon)$. The same closed path $C_\varepsilon$ generally maps to a loop $C_\omega$ enclosing spectral area on the complex energy plane. This property naturally allows one to define an eigenvalue winding number (EWN) of the spectral topology as

$$\frac{1}{2\pi i} \oint_{C_\varepsilon} d\vec{\varepsilon} \cdot \nabla_\varepsilon \ln \det(H(\varepsilon) - \omega_r I) \,, \tag{5}$$

where $\omega_r$ is an arbitrary reference energy and $I$ is an identity matrix with the same dimension as $H(\varepsilon)$. Alternatively, the EWN can be interpreted as characterizing the winding of a single eigenvalue on the complex-eigenvalue manifold $\mathcal{M}_\omega(\varepsilon)$ until $C_\omega$ forms a closed loop. For an $EP_2$, this requires the corresponding $C_\varepsilon$ to complete two cycles of winding around the $EP_2$. Eigenvalue winding behaviors have been experimentally observed in optical systems within a synthetic dimension [159]. Moreover, recent studies [161–163] have shown that braid and knot theories provide an alternative and powerful framework for characterizing the winding and braiding of multiple complex eigenvalues.

The eigenvectors of non-Hermitian Hamiltonians and their evolutionary behaviors are also drastically different from those of their Hermitian counterparts. Initially, researchers suggested that adiabatically varying the system parameters would allow the accumulation and observation of a Berry phase [14,164]. However, in non-Hermitian systems, the conventional adiabatic theorem breaks down and a nonadiabatic transition (NAT) is unavoidable [see the upper right panel of Fig. 4(a)] when circling the EP [165]. Consequently, slowly encircling an EP in the parameter space leads to chiral state transfer (CST) [37,38,152,166–174,39,40,43–45,123,150,151] where the final output state depends solely on the direction of encirclement, independent of the initial state. This unique behavior holds fundamental significance for the development of novel (quantum) optical devices and technologies [24,43,151,152,170,175–177]. For example, based on this phenomenon, an omnipolarizer can be realized in which the output light is polarized along a specific direction irrespective of the polarization of the input state [see the lower panel of Fig. 4(a)]. Later studies [40,178] revealed that CST can also occur in slowly varying non-Hermitian Hamiltonians along trajectories that do not encircle an EP [see Fig. 4(b)], sparking broad interest from both theory and experiment. Nevertheless, experimental realizations of EP-encircling chiral responses still face long-term challenges, including low transmission efficiency and large device footprints [39,168]. To address these issues, experiments have demonstrated that by employing moving EPs [150] [see Fig. 4(c)], the required parameter variations can be achieved adiabatically along a smaller encircling loop, leading to significant structural miniaturization and a remarkable improvement in mode transmittance. Almost the same time, CST has also been realized along open trajectories [151,152,170]. By carefully designing the evolution path, it is possible to maintain a low eigenstate evolution rate even when the system parameters undergo large variations or abrupt jumps [see Fig. 4(d)]. This enables the system to traverse rapidly while effectively bypassing path-dependent losses, thereby boosting the transmission efficiency in the designated direction to nearly unity in principle and further simplifying the evolution path [152]. The demonstrated trajectory can even be reduced to a straight line with an additional single point in the parameter space [see Fig. 4(e)]. In addition, dynamically encircling multiple or higher-order EPs has also sparked intense interest. Figure 4(f) illustrates that in multistate non-Hermitian systems, dynamically encircling arbitrary numbers of EPs gives rise to a robust chiral state-switching behavior, originating from multiple NATs among the eigenstates.

## B. Platforms for realizing nonlinearity–EP Coupling

Nonlinearity, where the system parameter depends on the field intensity, is ubiquitous feature of physical systems. It gives rise to rich phenomena such as harmonic generation, self-focusing, and solitonic propagation[179] that have no counterpart in purely linear systems. In recent years, tremendous attention has been devoted to the interplay between nonlinearity and non-Hermitian physics associated with EPs. Including the intensity dependence of system parameters not only yields a more accurate description for most realistic systems but also paves the way toward the next generation of reconfigurable EP-related devices, offering new opportunities for exploring exotic physical effects.

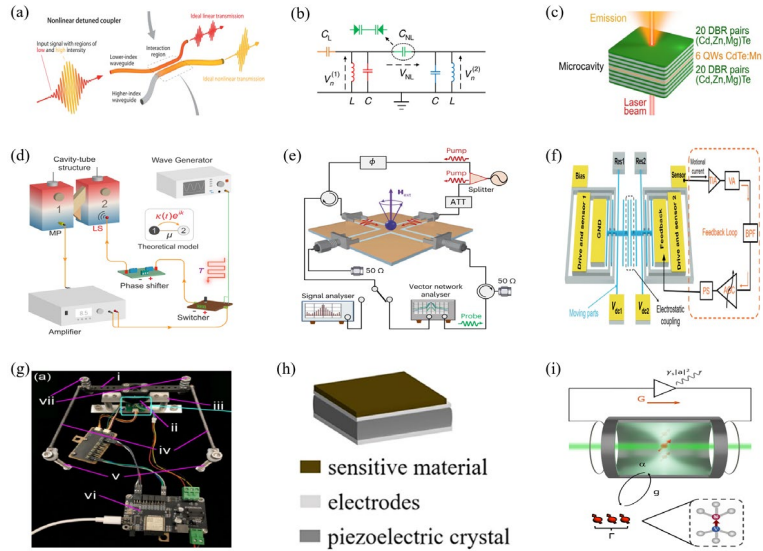

FIG. 5. Representative implementations of nonlinear non-Hermitian platforms. (a) Sketch of a nonlinear detuned coupler [180]. (b) A nonlinear circuit array [55]. (c) An exciton–polariton system[130]. (d-i) With the development of active control, more concise nonlinear coupling forms can be realized in (d) acoustics [61], (e) coupled PIM–KM platforms [132], (f) micromechanical resonator platforms[181], (g) mechanical systems [60] and (h) piezoelectric resonator platforms[102]. (f) In addition, nonlinearity have also been introduced in hybrid quantum systems [182]. The upper subpanel in panel (a) is reproduced with permission from Doppler et al., Nature 537(7618), 76–79 (2016). Copyright 2016 Springer Nature. The lower subpanel in panel (a) is reproduced with permission from Hassan et al., Phys. Rev. Lett. 118(9), 093002 (2017). Copyright 2017 American Physical Society. (b) Reproduced with permission from Nasari et al., Nature 605(7909), 256–261 (2022). Copyright 2022 Springer Nature. (c) Reproduced with permission from Liu et al., Phys. Rev. Lett. 124(15), 153903 (2020). Copyright 2020 American Physical Society. (d) Reproduced with permission from Li et al., Phys. Rev. Lett. 125(18), 187403 (2020). Copyright 2020 American Physical Society. (e) Reproduced from Shu et al., Light Sci. Appl. 13(1), 65 (2024), licensed under the Creative Commons Attribution (CC BY) license. (f) Reproduced with permission from Yu et al., Phys. Rev. Lett. 127(25), 253901 (2021). Copyright 2021 American Physical Society.

To this end, various nonlinear physical platforms have been demonstrated, as illustrated in Fig. 5. The implementation of nonlinearity in optical materials [see Fig. 5(a)] can be achieved through several distinct physical mechanisms. Fundamentally, nonlinear optics focuses on effective light–light interactions that can be described as a power series of the applied electric field involved through the high order susceptibility of materials. A prominent example is the Kerr effect [65,180] arising from the third-order susceptibility, where the refractive index becomes intensity dependent. A comprehensive overview of the relevant optical materials and device architectures can be found in this Review [183]. In active semiconductors, such as InGaAsP, a qualitatively different class of nonlinearities emerges from pump-controlled carrier population dynamics. In the unpumped or weakly pumped regime, most electrons reside in the valence band, and signal-induced inter band transitions result in intrinsic absorption that saturates with increasing optical intensity. Under stronger pumping, a substantial carrier population is excited to the conduction band, and radiative recombination with valence-band holes produces optical gain, which likewise saturates as the carrier reservoir is depleted. These carrier-mediated processes lead to saturable absorption and saturable amplification [80,184]. More recently, advances in inverse design and machine-learning–assisted photonic engineering [87,185,186] have enabled spatially resolved control over carrier distributions and material compositions, allowing nonlinear responses to be sculpted within two-dimensional geometries. Such artificially engineered nonlinearities substantially expand the accessible design space of optical materials, enabling programmable and reconfigurable forms of nonlinearity. These mechanisms can be naturally incorporated into the tight-binding Hamiltonian of Eq. (2), where each term may acquire a nonlinear form.

In electronic circuits [55,94,109] [Fig. 5(b)], early realizations were predominantly based on passive electrical components. More recently, these systems have evolved to incorporate active elements, including operational amplifiers and analog multipliers, which can be combined to implement negative-impedance converters, complex phase elements, high-frequency temporal modulation, and self-feedback mechanisms. A comprehensive overview of these contemporary developments and discusses their broader potential can be found in Ref. [187]. Nonlinearities in such circuits are typically introduced through nonlinear capacitive elements [55], nonlinear inductances [94,101], active components[107], or their combinations, all of which likewise allow nonlinear extensions of the Hamiltonian terms in Eq. (2).

Exciton polaritons[130,188–191] [Fig.5(c)] are hybrid light–matter quasiparticles formed under strong coupling between photons and excitons, the latter being collective electronic excitations corresponding to bound electron–hole pairs in semiconductors. Experimental observation of exciton polaritons requires strong light–matter interactions and a continuous optical pump to compensate for the inevitable loss of quasiparticles. Coupled with strong nonlinearity due to inherent quasiparticle interactions, this system can provide an alternative platform for studying the rich physics of nonlinear non-Hermitian systems. In the mean-field regime, exciton–polaritons can be treated as a scalar complex classical field governed by a driven–dissipative Gross–Pitaevskii equation with Kerr-type nonlinearity and gain saturation. Recent experimental advances have demonstrated the loading of exciton polaritons into a flat band of the Lieb lattice [192]; the realization of a PT-symmetric exciton-polariton condensate in coupled micropillars has been theoretically proposed [193]; and topological exciton-polariton insulators have also been predicted in two-dimensional exciton-polariton systems [194,195] and have been observed subsquently[196]. More recently, spectral winding topology and non-reciprocal behavior have also been reported in exciton–polariton platforms through geometrical twisting[190]. A recent Review[189] provides a comprehensive account of theoretical and experimental advances in room-temperature exciton–polaritons, together with an overview of the rich physical phenomena observed in these systems.

With the development of active control techniques, more concise realizations of nonlinear couplings have also been achieved in acoustics [59,61,197] [Fig. 5(d)], and coupled PIM–KM platforms [132] [Fig. 5(e)], micromechanical systems[181,198–200] [Fig. 5(f)], mechanical systems [60,201–203][Fig. 5(g)], and piezoelectric resonator platforms[102] [Fig. 5(h)], which can again be mapped onto nonlinear forms of the tight-binding model. For instance, in acoustic systems, an external feedback circuit—comprising a microphone, amplifier, wave-generator–driven switcher, phase shifter, and loudspeaker—can be employed. By precisely controlling the loudspeaker's amplitude and phase according to the signal detected by the microphone, this setup enables time-varying, unidirectional complex couplings, as well as the required acoustic loss or gain. A similar strategy can be applied in mechanical systems, active control enables tunable coupling, frequency adjustment, and programmable gain or loss for individual resonators. Such approaches provide flexible means to realize effective nonlinear coupling terms in the tight-binding Hamiltonian of Eq. (2), thereby greatly broadening the landscape of accessible non-Hermitian dynamics. Furthermore, nonlinear saturable gain has also been introduced hybrid quantum systems [58,182,204,205] [Fig. 5(i)], providing yet another versatile route to explore non-Hermitian nonlinear physics. More comprehensive discussions on nonlinear control and implementations can be found in dedicated review articles [145,183,204,206,207].

## III. FLEXIBLE SYSTEM CONTROL VIA NONLINEARITY

In nonlinear systems, the system parameters depend on the intensity. A direct consequence is that one can tune these parameters simply by controlling the intensity. To begin, we take nonlinear optics as an illustrative example. In Kerr nonlinear systems [66,67,208], one can introduce the pumping beam and signal beam. In general, nonlinearity is noninstantaneous, with a response time that is much longer than the phase-variation timescale of the pumping beam and signal beam[209]. Consequently, the system parameters governing the signal beam (which is first assumed to have with low intensity) can be directly tuned by adjusting the intensity of the pump beam. In a cavity quantum electrodynamics system[210], an incoherent pump field enables switching between the coherent perfect absorption regime and the near-perfect reflection regime through control of the linear absorption or gain. In addition to using a pump beam to control the signal, numerous studies have demonstrated that the signal itself can be tuned through its own intensity, leading to phenomena such as nonlinearity-induced topological phase transitions [55,56,67–69,180,211]. In addition to Kerr nonlinearity, materials with gain typically exhibit gain saturation, whereby the gain coefficient decreases once the optical intensity exceeds a certain threshold. This effect can be exploited to flexibly tune the effective nonlinearity strength. For instance, in PT symmetric systems that require balanced gain and loss, the stability condition enforces the gain to automatically match the fixed loss [107,115,116], thereby enabling self-adjustment through gain saturation. Therefore, it is natural to envision exploiting nonlinearity to enhance the tunability of the system, thereby realizing EPs (also possibly nonlinear version of EPs) and enabling the observation of their associated properties [59,67,68,71,72,212].

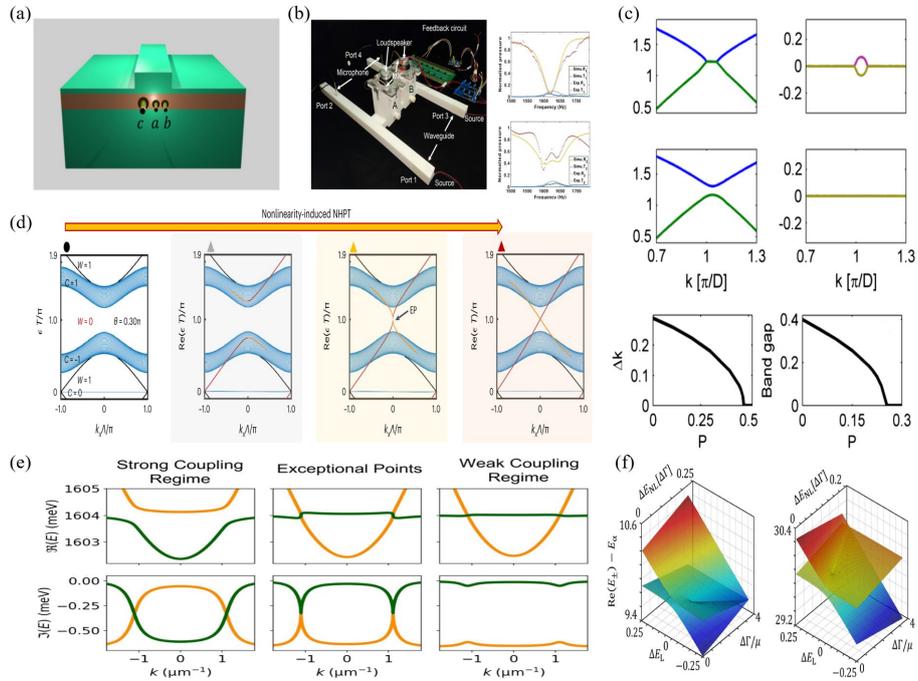

FIG. 6. Nonlinear control of system parameters for flexible observation of EPs and associated topological phenomena. (a) Nonlinear propagation of a pump beam $c$ and two probe beams $a$ and $b$ in a waveguide [212]. (b) Photograph of the experimental setup, where nonlinear feedback is employed to control loss and realize coherent perfect acoustic absorption at an EP [59]. (c) Real (left, upper panel) and imaginary (right, upper panel) components of the nonlinear spectrum, together with the size of the complex region (left, lower panel) and the gap of the real spectrum (right, lower panel) as functions of nonlinear power [72]. (d) Band structures as a function of nonlinearity, illustrating the nonlinearity-induced non-Hermitian phase transition and the emergence of an EP [67]. (e) Real (top) and imaginary (bottom) parts of the eigenvalues of a exciton–polariton system, illustrating the transition from spectra characteristic of strong coupling to those of weak coupling as the resonant pumping intensity is increased [130]. (f) Nonlinearity-induced rotation of the branch cut of the Riemann surfaces and shift of the EP [213]. (a) Reproduced with permission from Miri et al., New J. Phys. 18(6), 065001 (2016), licensed under the Creative Commons Attribution (CC BY) license. (b) Reproduced with permission from Xia et al., Phys. Rev. Lett. 135(6), 067001 (2025). Copyright 2025 American Physical Society. (c) Reproduced with permission from Lumer et al., Phys. Rev. Lett. 111(26), 263901 (2013). Copyright 2013 American Physical Society. (d) Reproduced with permission from Dai et al., Nat. Phys. 20(1), 101–108 (2024). Copyright 2024 Springer Nature. (e) Reproduced with permission from Opala et al., Optica 10(8), 1111-1117 (2023), licensed under the Creative Commons Attribution (CC BY) license. (f) Reproduced with permission from Wingenbach et al., Phys. Rev. Res. 6(1), 013148 (2024), licensed under the Creative Commons Attribution (CC BY) license.

In a waveguide structure with a second order nonlinearity [see Fig. 6 (a)], the conventional three-wave mixing process can be employed to explore EP physics [212]. In this configuration, the pump beam $c$ plays a role analogous to effective gain and loss, thereby establishing the conditions of PT symmetry for the two lower-frequency mixed beams $a$ and $b$. This provides a natural platform to investigate EPs. In an acoustic non-Hermitian scattering system composed of two-channel waveguides coupled to imbalanced lossy resonant cavities [see Fig. 6(b)], feedback control is implemented via an active acoustic unit to achieve accurate and decoupled tuning of non-Hermiticity, thereby realizing coherent perfect acoustic absorption at an EP [59]. In periodic PT photonic systems [72], Fig. 6(c) shows that the nonlinear spectrum varies with magnitude, where nonlinearity can transform the system from broken to full PT symmetry and vice versa. In a nonlinear non-Hermitian topological Su–Schrieffer–Heeger model based on coupled optical waveguides [68], nonlinearity governs the gain and loss, enabling active tuning of PT symmetry and the associated topological states. In a silicon nanophotonic Floquet topological insulator [see Fig. 6(d)], optical nonlinearity has emerged as a powerful tool for ultrafast manipulation of the non-Hermitian edge modes. As a result, these edge modes undergo dynamic phase transitions involving EPs, operating on timescales of only a few hundred picoseconds. Moreover, in exciton–polariton systems, the photonic and excitonic modes are coupled and subject to mode-dependent gain and loss. In the weak-pumping regime, these modes are strongly coupled via vacuum Rabi interaction, giving rise to hybridized polariton modes. As the pumping intensity increases, interactions with the light-induced incoherent particle reservoir drive a transition from the strong- to the weak-coupling regime, thereby enabling the emergence and observation of EPs [see Fig. 6(e)]. Furthermore, in this system, both the branch structure of the associated Riemann surfaces and the positions of the EPs can be continuously tuned by adjusting the strength of the polariton–polariton interaction as well as the polariton–reservoir interaction [see Fig. 6(f)]. In short, the introduction of nonlinearity not only greatly enriches the means of flexible control over the system, but also stimulates broader efforts that are expected to yield richer phenomena.

## IV. BEYOND LINEAR EP LIMITATIONS VIA NONLINEARITY

Despite their rich physical phenomena and promising applications, linear EPs still face severe challenges in realistic noisy environments, particularly in the long-standing debate on whether EP-based sensors indeed offer substantial advantages for detecting weak signals [10,32,140,214–219]. On the one hand, realizing EPs—especially higher-order EPs—requires highly complex experimental architectures with numerous, stable, and precisely tuned parameters [32,220,221]. As the order of the EP increases, the number of required tunable parameters grows rapidly. For instance, realizing an $EP_3$ already requires at least 6 independent tuning parameters. In addition, near an EP, the strong nonlinear dependence of eigenvalues and eigenstates on system parameters makes applications extremely sensitive to parameter variations, and thus the parameters must be adjusted with increasing precision. Such stringent parameter requirements therefore constitute a major bottleneck for the realization of EPs, especially higher-order ones.

On the other hand, ubiquitous noise severely hampers EP-related phenomena and applications. In realistic systems, noise arising from intrinsic material or imperfections is described as variations in Hamiltonian matrix elements [221,222], while noise from environmental interactions is modeled by stochastic Langevin terms [97,222]. The former depends on fabrication, material, and design—for example, fluctuations in capacitance due to Brownian motion of charges in circuit capacitors, or gain fluctuations in optical cavities due to unstable impurity populations in stimulated emission. In contrast, the latter is intrinsic to non-Hermitian systems, as the dynamics of such systems are generically accompanied by noise. Furthermore, this issue is particularly relevant in quantum settings, where the effects of vacuum noise cannot be ignored, especially when the dynamics involve amplification processes. The effects of such noise can be systematically analyzed using input–output theory and are strongly influenced by the nonorthogonality of the system's eigenvectors. As system parameters approach an EP, eigenvectors become nearly identical, their nonorthogonality diverges, and consequently the impact of system noise is greatly amplified. This amplification poses a fundamental obstacle to EP-based applications. Among them, sensing applications, which typically require operation in the immediate vicinity of an EP, are the most susceptible to noise. As a result, EP-based sensors naturally serve as the primary testbed for evaluating and quantifying the effects of noise in non-Hermitian systems. Based on a specific sensing scheme [219], it has been shown that EP-based sensors do not exhibit superior scaling behavior in terms of SNR. To establish a measurement-independent theoretical bound on SNR, the quantum Cramér–Rao bound formulated in terms of the quantum Fisher information (QFI) is commonly employed. Analyses based on the QFI formalism further indicate that linear EP-based sensors do not provide an intrinsic enhancement of SNR [217,218]. By contrast, in PT-symmetric EP sensors—specifically in coupled-cavity systems where the lasing transition coincides with an EP—the SNR was argued to be enhanced [140]. However, the validity of this conclusion has been questioned, as it relies on a linearized analysis that may not adequately capture the large fluctuations near the EP [217]. Using a self-consistent theory of fundamental noise in PT-symmetric EP sensors, subsequent studies [215] have shown that such sensors offer no advantage in terms of SNR once fundamental noise sources are taken into account. Beyond these theoretical investigations, the first experiment probing the quantum limits of EP-based sensors was performed using Brillouin laser gyroscopes [214]. The key observation was that the diverging enhancement of frequency splitting near the EP is precisely compensated by a simultaneous divergence in the laser linewidth. Although the present experiment examines only a specific physical platform, it clearly demonstrates that, at least within this class of systems, the SNR is not improved for EP-based sensors.

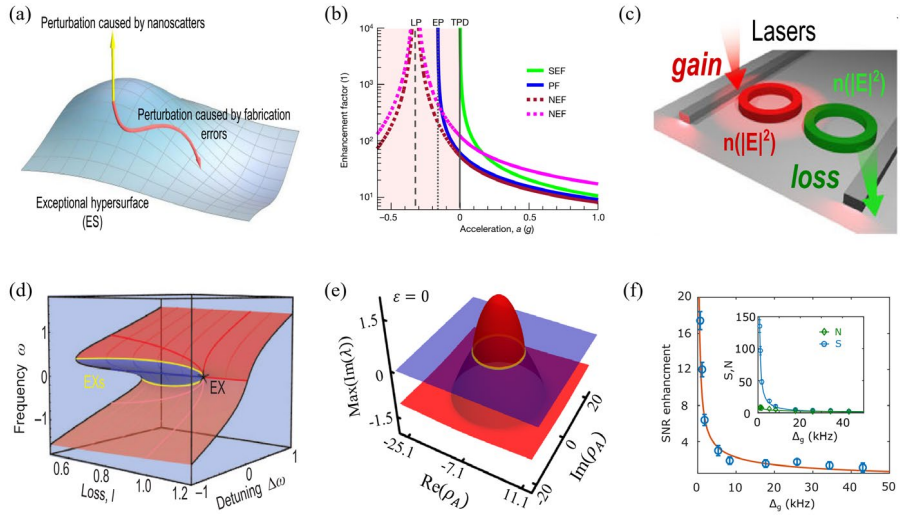

FIG. 7. Recent advances in addressing parameter-tuning complexity and noise divergence near EPs. (a) Exceptional hypersurfaces [31] can combine robustness with sensitivity, effectively reducing parameter dependence along specific directions. (b) Sensitivity enhancement factor (green line), Petermann factor (blue line), and noise enhancement factors as functions of the applied acceleration [222]. (c) Schematic of two coupled cavities with nonlinear gain and nonlinear loss [101], illustrating strong suppression of noise enhancement. (d) Steady-state eigenfrequencies of two coupled cavities, one with saturable gain and the other with linear loss [94]. The red and blue regions represent stable and unstable states, respectively. (e) Maximum imaginary part of the nonlinear Hamiltonian eigenvalues as a function of time-dependent oscillatory components, showing that noise-induced fluctuations are suppressed by a hidden feedback mechanism and remain finite [223]. (f) Experimental demonstration in a realistic hybrid quantum system [182], where the signal-to-noise ratio (SNR) is enhanced by a factor of 16. (a) Reproduced with permission from Zhong et al., Phys. Rev. Lett. 122(15), 153902 (2019).

In linear systems, singular structures such as EP lines [16,125,224], surfaces [225–227], and spheres [228] offer two major advantages over isolated EPs. On the one hand, they introduce anisotropic dependencies in the responses of eigenvalues and eigenstates, which facilitates applications based on anisotropic effects [229]. On the other hand, they can partially relax the stringent parameter requirements for forming EPs, thereby enhancing the robustness of EP-related applications [31,230]. Figure 7(a) shows a hypersurface of EPs that combines robustness and sensitivity. Undesired perturbations arising from fabrication imperfections or experimental uncertainties shift the spectrum along the hypersurface, thereby keeping the system on the EP hypersurface. In contrast, perturbations associated with the measured quantities—for example, those induced by nanoscatterers—drive the spectrum out of the surface, i.e., away from the EP hypersurface. To address the enhanced noise effects associated with eigenmode coalescence, sensors operating at the transmission peak degeneracy (TPD) have been experimentally demonstrated [see Fig. 7(b)], yielding a threefold enhancement in signal-to-noise ratio (SNR) compared with configurations operating away from the TPD.

With increasing efforts devoted to nonlinear non-Hermitian systems, it has been shown that nonlinearity-induced feedback mechanisms can suppress noise, thereby further enhancing the SNR. Figure 7(c) illustrates two coupled resonators with saturable gain and saturable loss, where the experimental setup demonstrates a two-order SNR enhancement in voltage variation measurements. A more efficient configuration is achieved when only one cavity incorporates saturable gain while the other provides linear loss. Figure 7(d) shows the corresponding steady-state eigenfrequencies, where the red and blue regions denote stable and unstable states, respectively. The yellow lines mark $EP_2$ lines at the boundaries of regions with different stabilities, and their intersection gives rise to an EX, i.e., an EP$_3$. Compared with linear EP$_3$, the number of tunable real parameters in the Hamiltonian matrix is reduced from six to two, significantly alleviating the parameter-tuning complexity for realizing higher-order EPs in experiments. In addition, noise is strongly suppressed, resulting in a substantial improvement in SNR. Away from the EX, the effect of noise can be quantified by a linear approximation[97]. To rigorously quantify the effect of noise on SNR without approximation, Fig. 7(e) shows the maximum imaginary part of the Hamiltonian eigenvalues as a function of fluctuation, which effectively characterizes the dynamical process of noise-induced fluctuations. Since the imaginary part remains below zero for large fluctuations, the fluctuations inevitably decay during evolution, ensuring their finiteness through a hidden feedback mechanism. This provides a strict theoretical guarantee for a SNR enhancement. Experiments on an NV–Van der Pol hybrid system, which exhibit a similar nonlinear non-Hermitian Hamiltonian, confirm a 16-fold improvement in SNR [see Fig. 7(f)].

## V. NEW SINGULARITIES AND MECHANISMS

Singularities are critical points at which the behavior of a model governing a physical system undergoes a fundamental change compared with that at neighboring regions. They ubiquitously exist across disciplines and play a pivotal role both in probing fundamental physical laws and in enabling a broad range of applications. Prominent examples include two-dimensional Dirac points in graphene [231–233], Weyl points in topological semimetals [234,235], bound states in the continuum in photonics [236,237], van Hove singularities[238] in condensed matter systems, among others [239].

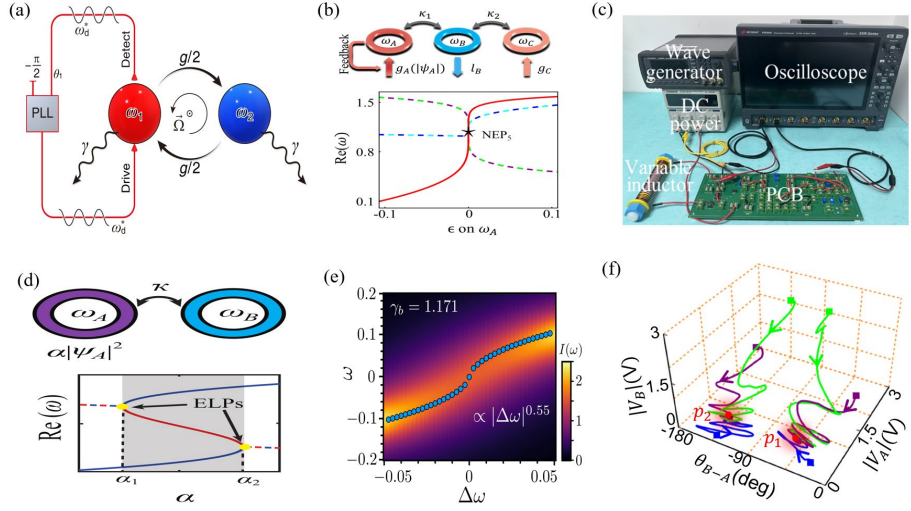

FIG. 8 New singularities and their mechanisms in nonlinear non-Hermitian systems. (a) Higher-order singularities in phase-tracked dynamics [93]. Mode 1, driven by an external force, is coherently coupled to Mode 2, while a phase-locked loop (PLL) enables closed-loop phase-tracked oscillations. (b) The upper panel shows a schematic of the realization of nonlinear exceptional points (NEPs) with a complete basis in dynamics [104], while the lower panel presents the eigenvalue evolution versus external perturbation, revealing the coalescence of the stable mode with four auxiliary modes. (c) The experimental setup implementing a minimal scheme for observing a NEP$_3$ [98]; (d) The upper panel shows a schematic of two coupled resonators, where the resonance frequency of the left resonator depends on the wave amplitude via a Kerr-type nonlinearity, while the lower panel depicts the real part of the eigenvalues as the Kerr coefficient varies, revealing the existence of EP-like points [105]. (e) Noise–nonlinearity interplay that induces an EP$_2$ at a new position in parameter space under strong noise conditions [97]. (f) Measured evolution trajectories in phase space with different initial states. Due to the attractor nature of the stable states, chiral-like state transfer between steady states can be realized by hopping through three distinct points in parameter space [118]. (a) Reproduced from Zhou et al., Nat. Commun. 14(1), 7944 (2023), licensed under the Creative Commons Attribution (CC BY) license. (b) Reproduced with permission from Bai et al., Phys. Rev. Lett. 130(26), 266901 (2023). Copyright 2023 American Physical Society. (c) Reproduced with permission from Bai et al., Phys. Rev. Lett. 132(7), 073802 (2024). Copyright 2024 American Physical Society. (d) Reproduced with permission from Fang et al., Phys. Rev. B 111(16), L161102 (2025). Copyright 2025 American Physical Society. (e)

In the emerging domain of nonlinear non-Hermitian systems, a variety of new singularities and associated physical phenomena have recently been discovered. In phase-tracked electromechanical oscillators [93], a previously unexplored third-order singularity has been demonstrated both theoretically and experimentally in a pair of coherently coupled mechanical modes [Fig. 8(a)], where mode 1 is linearly excited by an external sinusoidal force and a phase-locked loop (PLL) is employed to realize closed-loop phase-tracked oscillations. Similarly, an external measurement-feedback loop [240] enables remarkable linewidth narrowing and enhanced measurement sensitivity, thereby greatly extending the scope of potential applications. In addition, leveraging the intrinsic nonlinearity of materials not only simplifies experimental complexity but also gives rise to novel singularities and intriguing phenomena. The upper panel of Fig. 8(b) illustrates a model composed of a nonlinear saturable-gain cavity coupled with a linear-loss and a linear-gain cavity (a gain model with a much larger saturation threshold). This nonlinear system exhibits a fifth-order nonlinear EP (NEP$_5$), at which one stable self-consistent eigenstate and four auxiliary self-consistent eigenstates of the nonlinear Hamiltonian coalesce, in direct analogy to the eigenstate coalescence observed at a linear EP. The lower panel of Fig. 8(b) shows the real part of the eigenfrequencies, where the coalescence of these five self-consistent eigenstates ensures that NEP$_5$ retains the essential features of a linear EP. Meanwhile, the system dynamics are governed by a $3 \times 3$ instantaneous Hamiltonian, whose eigenstates are generally distinct from the five self-consistent eigenstates of the nonlinear Hamiltonian. As a consequence, the eigenstates of the instantaneous Hamiltonian do not necessarily coalesce at the NEP$_5$. In addition, compared with realizing a linear EP$_5$, the number of parameters requiring precise tuning is drastically reduced. Therefore, NEPs are particularly well suited for a broad range of EP-related applications. Figure 8(c) shows an experimental setup implementing a minimal scheme that combines non-Hermitian coupling with nonlinear saturable gain. The circuit consists of two LC resonators coupled through a capacitor and a resistor, which together provide the non-Hermitian coupling. One of the resonators incorporates a saturable gain realized by a linear voltage amplifier and two diodes. In this circuit, a $NEP_3$ with a complete basis in dynamics has been successfully observed.

Interestingly, even in the absence of non-conservative elements such as gain, loss, or nonreciprocal hopping, exceptional features can also arise in nonlinear Hermitian systems. The upper panel of Fig. 8(d) shows a schematic of two coupled resonators, where the resonance frequency of the left resonator depends on the wave amplitude through a Kerr-type nonlinearity. This system hosts new singularities, termed EP-like points (ELPs), whose critical behavior mirrors that of conventional EPs [see the lower panel of Fig. 8(d)]. The interplay between noise and nonlinearity can also induce new type of singularities. Figure 8(e) shows the peak frequency and power spectrum under strong Gaussian white noise, where an $EP_2$ emerges at a new position in parameter space. With the deepening exploration of nonlinear systems, it has become clear that the dynamics around nonlinear singularities are considerably richer. For instance, encirclement around an ELP in a system without non-Hermitian elements can lead to multifrequency states. In systems with saturable gain, different initial states can be attracted to a stable mode (if it exists), reflecting the attractor nature of nonlinear steady states [see Fig. 8(f)]. Consequently, when encircling a $NEP_3$, the chiral phenomena strictly arise from the basins of attraction of stable states and are not constrained by adiabaticity. As a result, the parameter trajectory does not necessarily need to include the $NEP_3$. Remarkably, it can even be simplified to three distinct points by varying only a single parameter in the parameter space.

## VI. PHENOMENA AND APPLICATIONS VIA NONLINEARITY–EP COMBINATION

Various novel physical phenomena and potential applications arising from the synergy between nonlinearity and EPs are rapidly emerging, which will further stimulate the prosperity of this field. Beyond fundamental demonstrations, these advances pave the way toward practical implementations—ranging from robust wireless power transfer [Fig. 9(a)] and enhanced frequency-comb generation [Fig. 9 (b)] to exceptional state transfer in lasers [Fig. 9 (c)], on-chip optical isolators [Fig. 9 (c, d)], and ultra-efficient nonlinear optical signal processing [Fig. 9 (f)], with further examples continuously being uncovered.

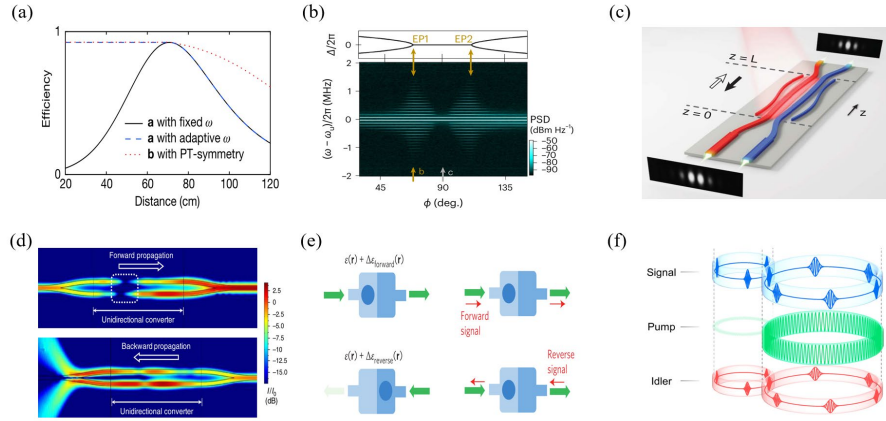

FIG. 9. Novel phenomena and applications enabled by the combination of nonlinearity and EPs. (a) Power transmission efficiency as functions of the coupling coefficient, demonstrating robust efficiency wireless power transfer [107]. (b) Upper panel: evolution of the real parts of two eigenfrequencies; lower panel: measured radiation spectra, showing clear enhancement of magnonic frequency combs near the two EPs [132]. (c) On-chip mode-locked laser realized through exceptional state transfer, enabling the formation of a single topological mode independent of the path in parameter space [167]. (d) Optical intensity distributions in a coupled-waveguide structure supporting EP encircling under gain saturation, leading to ultrabroadband, on-chip optical nonreciprocity [44]. (e) Schematics of reciprocity analysis for small additional waves in the presence of a large forward signal, confirming reciprocity for the additional waves [241]. (f) Schematic of the intracavity fields in PT-symmetric coupled microresonators, where the signal, pump, and idler distributions are shown in blue, green, and red, respectively. Near the EP, the system enables ultra-efficient nonlinear optical signal processing [242]. (a) Reproduced from Assawaworrarit et al., Nature 546(7658), 387–390 (2017). Copyright 2017 Springer Nature. (b) Reproduced from Wang et al., Nat. Phys. 20(7), 1139–1144 (2024). Copyright 2024 Springer Nature. (c) Reproduced from Schumer et al., Science 375(6583), 884–888 (2022). Copyright 2022 American Association for the Advancement of Science. (d) Reproduced from Choi et al., Nat. Commun. **8**(1), 14154 (2017), licensed under the Creative Commons Attribution (CC BY) license. (e) Reproduced from Shi et al., Nat. Photonics **9**(6), 388–392 (2015). Copyright 2015 Springer Nature. (f) Reproduced

Wireless power transfer has paved the way toward real-world applications such as wireless powering of implantable medical devices and wireless charging of stationary electric vehicles. To ensure robust efficiency, a PT-symmetric circuit composed of two LC resonators—one of which incorporates a nonlinear gain-saturation element—was demonstrated, enabling transfer efficiency to remain nearly unity over distance variations of almost one meter without any active tuning [see Fig. 9(a)] [107]. Subsequently, frequency-stable robust wireless power transfer was realized [109], benefiting from the fact that the nonlinear stable state eigenfrequencies remain almost unchanged with respect to coupling variations when operating away from higher-order EPs. More recently, communication-free robust wireless power transfer with constant output power and stable frequency [113], as well as one-transmitter–multiple-receiver schemes exploiting exceptional-point degeneracies [111], have also been demonstrated, reflecting the flourishing development of this research area.

A frequency comb [243] is a spectrum of equally spaced discrete frequency components with high time–frequency accuracy, making it indispensable for applications in precision spectroscopy [244], satellite navigation [245], and atomic clocks [246]. Conventional generation methods rely on weak material nonlinearities and require high power densities, which limits their practicality. By exploiting EPs in the nonlinear coupling [132] between pump-induced magnon mode (PIM) and intrinsic Kittel mode (KM), magnonic frequency combs can be efficiently produced at low pump powers with excellent tunability. As shown in Figure Fig. 9(b), the $EP_2$s are constructed by the precise control of the PIM-KM coupling strength through the inherited chirality of PIM induced by a magnetic field. The probe simultaneously perturbs both the PIM and KM, thereby facilitating their nonlinear coupling and enabling frequency-comb generation. This mechanism remains robust across a wide range of pump powers and sample geometries. As the system is tuned toward an EP, the efficiency of magnonic frequency-comb generation is consistently enhanced, with the number of comb teeth increasing markedly from only a few to as many as 32. Each comb tooth corresponds to a higher-order PIM, which potentially contributed to the construction of a PIM ensemble by customizing the profile of comb teeth. These results establish a new benchmark for frequency-comb generation.

Figure 9 (c) illustrates the concept of an EP-encircling laser [167]. The separation between the auxiliary waveguides and their respective main waveguides introduces detuning, while the separation between the two main waveguides provides coupling. By modulating structural parameters along the propagation direction, the dynamical encircling of the induced EP can be realized. The red waveguide incorporates saturable gain, favoring a spatially evolving mode that continuously transforms from one eigenstate profile to another while avoiding nonadiabatic transitions. Consequently, the spatially evolving mode faithfully settles into a pair of bi-orthogonal states at the two opposing facets of the laser cavity. Nonreciprocal light propagation is a central requirement for optical isolators and circulators, which play vital roles in signal processing, telecommunications, and the protection of high-power lasers. By combining gain-saturation nonlinearity with the non-Hermiticity associated with EPs, high-efficiency nonreciprocity can be realized [44,247]. As shown in Fig. 9(d), the coupled-waveguide structure is designed to dynamically encircle an EP along the propagation direction. Two-dimensional simulations reveal distinct modal intensity profiles for excitations from opposite directions, clearly manifesting the chiral state transfer phenomenon. In this structure, the introduction of saturable gain ensures that different steady-state modes correspond to different effective system parameters, thereby enabling robust nonreciprocal transport. Notably, the resulting nonreciprocal efficiency is independent of wavelength, allowing the effect to be realized across a broad spectral range. However, the realization of optical isolators is still fundamentally constrained by the so-called dynamic reciprocity [241]. Specifically, in the presence of a strong signal, any additional weak waves experience the dielectric distribution fixed by the dominant signal [see Fig. 9(e)]. As a result, the system behaves reciprocally for small perturbations, limiting the isolator's effectiveness.

Another interesting application of EPs is ultra-efficient nonlinear optical signal processing (NOSP) [242]. Conventional NOSP suffers from weak optical nonlinearities. A common strategy to enhance nonlinear effects is to employ high-quality-factor (Q) microresonators. However, the narrow resonance linewidth, inversely related to Q, leads to temporal mixing of information (intersymbol interference).

This results in a fundamental tradeoff between the enhanced nonlinearity provided by high-Q cavities and the maximum signal bandwidth, which scales with the data rate efficiency. To overcome this bandwidth–efficiency limit, a PT-symmetric microresonator system has been designed to simultaneously boost light intensity and enable high-speed operation. The key concept is the coexistence of a PT-symmetry-broken regime for the narrow-linewidth pump wave and near-EP operation for broadband signal and idler waves [Fig. 9(f)], thereby relieving the bandwidth–efficiency limit imposed on conventional single-resonator systems.

## VII. CONCLUSIONS AND OUTLOOK

In this Review, we have systematically summarized recent advances in the synergy between nonlinearity and EPs. Our focus has been primarily on classical wave systems, along with certain quantum systems whose dynamics are effectively described by non-Hermitian Hamiltonians rather than by Lindblad master equations or Heisenberg–Langevin approaches. We have discussed the experimental platforms that support diverse nonlinear effects, the emergence of novel singularities and mechanisms, as well as the associated physical phenomena and applications. These developments highlight the potential of nonlinear–EP systems to overcome the intrinsic limitations of linear EP physics and to stimulate further progress across photonics, acoustics and beyond.

Despite these advances, important challenges remain. For instance, EP sensing has been accompanied by continuous debate since its proposal [141]. Although experiments have demonstrated advantages in signal amplification and weak-signal detection, the improvement of the SNR in noisy environments remains highly controversial. In classical and hybrid quantum systems, certain works suggest that feedback mechanisms can significantly enhance the SNR, yet this topic is still under discussion. Moreover, both linear and nonlinear EP sensors require the signal to persist for a sufficient duration in order for the system to reach a stable state, and exploring ways to design and harness noise to reduce this stabilization time presents an exciting avenue for future research. In addition, how to strategically exploit EPs in fully quantum systems to improve sensor performance remains an open question [248].

Although the CST phenomenon has been demonstrated across various platforms, it still faces important limitations. In purely lossy systems, the transmission efficiency in one direction can approach unity, yet the opposite direction remains nearly zero. Achieving bidirectional high efficiency requires the incorporation of gain materials, which introduces additional challenges for device fabrication. In nonlinear systems, new CST mechanisms based on stable state attractors have been proposed and exemplarily verified in circuit platforms, but their applicability to realistic devices remains absence and may bring unforeseen issues. Furthermore, current schemes are largely restricted to the asymmetrical conversion of a pair of fundamental modes, which constrains the multiplexing capacity. A promising route forward is to explore novel EP-encircling optical structures that enable asymmetrical conversion of more complex field modes—such as multimode states, higher-order modes, polarization states, and orbital angular momentum—thereby significantly enhancing their potential for information processing [249]. In addition, EP-encircling lasers provide an appealing pathway to mode-locked lasing and hold considerable promise as on-chip light sources. Existing demonstrations, however, remain limited to locking low-order modes with two coupled waveguides. Employing multi-waveguide arrays that support single-mode output is expected to substantially increase lasing power while also enriching the functionalities of EP-encircling lasers.

Symmetry plays a central role and lies at the origin of many fundamental phenomena in complex systems across the natural sciences. In the presence of symmetries or generalized similarities, specific constraints impose among the matrix elements of the effective Hamiltonian, thereby reducing the number of independent parameters that must be tuned to reach EPs. As a consequence, the codimension of EPs can be lowered, which greatly facilitates the identification, classification, and exploration of higher-order EPs and their associated properties[250,251]. Recent studies have shown that[252] generalized similarities unify unitary and anti-unitary symmetries, providing a new framework for understanding the emergence of EPs in lower dimensions. Moreover, in nonlinear non-Hermitian systems, PT symmetry has played an eminent role in understanding EP structures and the associated phase transitions[253]. Despite this progress, the interplay of symmetry, nonlinearity, and non-Hermiticity continues to represent an open and actively developing frontier, with the influence of symmetry on an even richer variety of phenomena yet to be further explored.

At present, studies on nonlinear (non-)Hermitian exceptional structures are largely confined to zero-dimensional points. In contrast, richer classes of singular structures—such as one-dimensional rings and arcs, as well as higher-dimensional surfaces including tori, saddles, and spheres—remain relatively unexplored. These new types of singularities could provide unique opportunities for investigating anisotropic effects, as well as a wide range of other emergent phases of matter and novel physical phenomena. In linear non-Hermitian systems, it has been theoretically proposed and experimentally demonstrated that non-Hermitian perturbations can deform Dirac[13] or Weyl[224,254,255] points into exceptional rings, giving rise to new spectral and topological structures. These exceptional rings host rich topological characteristics and provide unique opportunities for exploring anisotropic responses, non-Hermitian topological transport, and other unconventional physical phenomena that are absent in Hermitian systems. More recently, exceptional rings[256] have also been investigated in nonlinear non-Hermitian planar optical microcavities, where their topological invariants and responsivity were shown to persist and acquire new features. These advances naturally motivate future efforts toward uncovering and classifying higher-dimensional nonlinear exceptional structures and understanding their roles in shaping non-Hermitian topology and dynamics

In addition, understanding and exploiting non-Hermiticity associated with EPs in nonlinear topological structures can open up numerous research opportunities for both fundamental studies and future applications. At present, a universal theoretical framework that concurrently integrates non-Hermiticity, topology, nonlinearity, and symmetries remains scarce. Building on catastrophe theory, the recently reported universal topology[96] of EPs in nonlinear non-Hermitian systems provides an important step in this direction. This framework offers a unifying perspective on the organization and classification of nonlinear EPs, and holds promise for guiding future experimental discoveries. Further development of such theoretical frameworks would provide both qualitative and quantitative descriptions of their mutual interplay, enabling precise and flexible control over the complex interactions governed by diverse mechanisms. As an illustrative example, in nonlinear non-Hermitian planar optical microcavities, the emergence of two concentric exceptional rings can be precisely and flexibly controlled through system parameters, and their topological invariants and responsivity have been investigated [256]. Nevertheless, a number of fundamental challenges and limitations remain. In particular, the number of nonlinear eigenmodes may change with system parameters, which can modify the dimensionality of the complex-eigenvalue manifold. This raises nontrivial questions regarding how eigenvalue and eigenstate topology—and the associated topological invariants—should be defined and characterized on such parameter-dependent manifolds. Moreover, whether and how the topological properties of nonlinear non-Hermitian systems influence their dynamical evolution remains largely unexplored. This, in turn, would advance the development of topological concepts in nonlinear non-Hermitian systems hosting EPs, and equip them with attributes such as topological robustness, exceptional sensitivity, remote controllability, and flexible reconfigurability.

## ACKNOWLEDGMENTS

This work is supported by the National Key Research and Development Program of China [Grant No. 2022YFA1404900], the National Natural Science Foundation of China [Grants No. 12274332, No. 12334015, 12321161645, 12404440], the China Postdoctoral Science Foundation under Grant No. 2023M742715, the China National Postdoctoral Program for Innovative Talents under Grant No. BX20240266 and Postdoctor Project of Hubei Province under Grant No. 2024HBBHCXB054.

## AUTHORDECLARATIONS

### Conflict of Interest

The authors have no conflicts to disclose.

### Author Contributions

**K. Bai:** Conceptualization (lead); Formal analysis (lead); Investigation (lead); Methodology (lead); Visualization (lead); Writing - original draft (lead); Writing - review & editing (equal). **C. Lin:** Conceptualization (supporting); Investigation (supporting); Writing - review & editing (supporting). **J. Li:** Conceptualization (supporting); Investigation (supporting); Writing - review & editing (supporting). **M. Xiao:** Conceptualization (equal); Funding acquisition (lead); Supervision (lead); Writing - review & editing (lead).

### DATAAVAILABILITY

Data sharing is not applicable to this article as no new data were created or analyzed in this study.